\documentclass[letterpaper, 10 pt, conference]{ieeeconf}  

\IEEEoverridecommandlockouts                              
\usepackage{graphics} 
\usepackage{epsfig} 
\usepackage{mathptmx} 
\usepackage{times} 
\usepackage{amsmath} 
\usepackage{amssymb}  
\usepackage{xcolor}
\usepackage{graphicx,subfigure}
\usepackage{cancel}

\newtheorem{remark}{Remark}[section]
\usepackage{graphicx}

\title{\LARGE \bf Active queue management: \\ First steps toward a new control-theoretic viewpoint}
 
\author{C\'{e}dric JOIN$^\text{a, e}$, Hugues MOUNIER$^\text{b}$, Emmanuel DELALEAU$^\text{c}$, Michel FLIESS$^\text{d, e}$ 
\thanks{$^\text{a}$CRAN, (CNRS, UMR 7039), Universit\'{e} de Lorraine, B.P. 70239, 54506 Vand{\oe}uvre-l\`{e}s-Nancy, France. \newline {\tt\small cedric.join@univ-lorraine.fr}
}
\thanks{$^\text{b}$L2S (UMR 8506), Universit\'{e} Paris-Saclay, CentraleSup\'elec, CNRS, 
3 rue Joliot-Curie, 91192 Gif-sur-Yvette, France. \newline {\tt\small hugues.mounier@universite-paris-saclay.fr}
}
\thanks{$^\text{c}$IRDL (CNRS, UMR 6027), \'Ecole nationale d'ing\'enieurs de Brest, 29200 Brest, France. \newline {\tt\small emmanuel.delaleau@enib.fr}
} 
\thanks{$^\text{d}$LIX (CNRS, UMR 7161), \'Ecole polytechnique, 91128 Palaiseau, France. {\tt\small Michel.Fliess@polytechnique.edu, \newline michel.fliess@swissknife.tech}
}
\thanks{$^\text{e}$AL.I.E.N., 7 rue Maurice Barr\`{e}s, 54330 V\'{e}zelise, France. \newline {\tt\small \{cedric.join, michel.fliess\}@alien-sas.com}
}
}

\begin{document}

\maketitle
\thispagestyle{empty}
\pagestyle{empty}

\begin{abstract}
\hspace {0.05cm}
Active Queue Management (AQM) for mitigating Internet congestion has been addressed via various feedback control syntheses, among which P, PI, and PID regulators are quite popular and often associated to a Smith predictor. Here, to better account for the many uncertainties, like the round trip time or the number of TCP sessions, 
the above controllers are replaced by an intelligent proportional controller associated to Model-Free Control (MFC) and by forecasting techniques derived from a new approach to time series. Several computer simulations via a well accepted linear modeling, where the delay is assumed to be constant, are presented and discussed.   

\textbf{\textsl{Keywords}}---{\hspace {0.05cm} 
Internet congestion, active queue management, model-free control, intelligent proportional control, delay, time series, nonstandard analysis, prediction.}

\end{abstract}

\section{Introduction}\label{intro}
The aim of \emph{Active Queue Management}, or \emph{AQM}, is to mitigate the disastrous effects of Internet congestion, when using the \emph{Transmission Control Protocol}, or \emph{TCP}. An abundant literature has been devoted to this hot topic since 30 years, where control theory often plays a key r\^{o}le (see, \textit{e.g.}, \cite{adams,varma}), especially perhaps P, PI, PD, and PID regulators: See, \textit{e.g.}, \cite{pie} for the popular \emph{PIE}, or {\em Proportional Integral Enhanced} controller. Like many control-oriented investigations until today we are also using a linear time-invariant delay system \cite{hollot1,hollot2}. It derives from an approximate linearization around an operating point of a much more complex nonlinear modeling \cite{misra,olu}.
Even with such a simplified transfer function, the parametric uncertainties and the presence of a delay render the dropping packets policy inside a buffer quite involved, in spite of most promising recent results (see, \textit{e.g.}, \cite{hotchi,hotchi1,hotchi2}).

This linear delay system is used here solely
for simulation purposes and not for AQM. \emph{Model-free control}, or \emph{MFC}, in the sense of \cite{mfc1,mfc2}, is introduced in this communication for AQM, \textit{i.e.}, for improving the adaptability to the dynamic nature of network traffics. The quite easy gain tuning and the remarkable robustness of the \emph{intelligent} controllers associated to MFC explains its worldwide popularity: See, \textit{e.g.}, the references in \cite{mfc1,mfc2}, and \cite{hydrogen,grenoble,li,liu,lv,montreal,olama,sun,zhang} for some most recent and promising applications to various concrete subjects. Note here that MFC has already been suggested for investigating more or less related questions on the Internet of Things \cite{iot} and cybersecurity \cite{sauter}.
\begin{remark}
Let us single out \emph{ramp metering} \cite{alinea,dirn}, \textit{i.e.}, another type of queue management in order to control the rate of vehicles entering highways. MFC provides a remarkably flexible and robust algorithm  which is used in Northern France.
\end{remark}

In the present work, the \emph{ultra-local model} \cite{mfc1} does not read 
\begin{equation}\label{1}
\frac{d}{dt}{\delta q}(t) = F(t) + \alpha \delta u(t)    
\end{equation}
but
\begin{equation}\label{2}
\frac{d}{dt}\delta q(t) =F(t) + \alpha \delta u(t - h)    
\end{equation}
where
\begin{itemize}
\item $\delta u$ and $\delta q$ are respectively the input and output variables, 
\item $F$ is a quantity which subsumes the poorly known system structure and the disturbances,
\item $\alpha$ is a constant such that the three terms in Equation~\eqref{1} are of the same order of magnitude,
\item $h >0$ is an unavoidable non-negligible  delay, which for simplicity's sake is assumed to be constant and known, like in all the numerous publications using a linear approximation.
\end{itemize}
For Equation \eqref{1}, an efficient real-time estimate $F_{\rm est}$ of $F$ has been obtained \cite{mfc1} via a data-driven closed-form formula. It yields a kind of feedback equivalence with the  integrator $\frac{d}{dt}\delta q(t) =  \alpha \delta v(t)$ where $\delta v(t) = \delta u(t) + \frac{F_{\rm est}(t)}{\alpha}$. Identical calculations  for Equation \eqref{2} yield
$
\frac{d}{dt}\delta q(t) =  \alpha \delta v(t - h)   
$,
where 
$$
\delta v(t - h) = \delta u(t - h) + \frac{F_{\rm est}(t)}{\alpha}
$$ 
Smith predictors \cite{smith,visioli}, which are prevailing in the model-based linear approach to AQM\footnote{However there are some exceptions: See, \textit{e.g.}, \cite{xu}.}, do not fit in our model-free context since there is no closed-form mathematical expression of $F$. We follow here a paper on \emph{supply chain management} \cite{hamiche} and replace Smith predictors by forecasting techniques borrowed from a new understanding of time series in financial engineering~\cite{perp,agadir}. This setting, which is based on a quite recent result~\cite{cartier} in \emph{nonstandard analysis}, has also been exploited in solar energy management \cite{solar} and cloud computing \cite{cloud}. It might lead in the future to some advances in an important control-theoretic question (see, \textit{e.g.}, \cite{plestan} for an excellent overview).

Our paper is organized as follows. The replacement of the Smith predictors by forecasting time series techniques is detailed in Section \ref{tool} as well as its application to model-free control. After presenting the linear model for the numerical simulations, Section \ref{simu} displays and discusses several scenarios where 
\begin{itemize}
    \item the round trip time and the number of TCP sessions may change;
    \item the presence of UDP flows (see, \textit{e.g.}, \cite{udp}) and other exogenous perturbations is examined.
\end{itemize}
Section \ref{conclu} suggests future investigations.

\section{Tools for replacing Smith predictors}\label{tool}
\subsection{Forecasting via time series}\label{basics}
\subsubsection{Time series}
Take the time
interval $[0, 1] \subset \mathbb{R}$ and introduce as often in
\emph{nonstandard analysis} (see \cite{robinson}, and \cite{diener,lobry}) the infinitesimal sampling ${\mathfrak{T}} = \{ 0 = t_0 < t_1 < \dots < t_\nu = 1 \}$
where $t_{i+1} - t_{i}$, $0 \leqslant i < \nu$, is {\em infinitesimal},
{\it i.e.}, ``very small''. A
\emph{time series} $X(t)$ is a function $X: {\mathfrak{T}}
\rightarrow \mathbb{R}$.

A time series ${\mathcal{X}}: {\mathfrak{T}} \rightarrow \mathbb{R}$
is said to be {\em quickly fluctuating}, or {\em oscillating}, if,
and only if, the integral $\int_A {\mathcal{X}} dm$ is
infinitesimal, for any \emph{appreciable} interval $A$, \textit{i.e.}, an interval which is neither ``very small'' nor ``very large.''

According to a theorem due to Cartier and Perrin~\cite{cartier} the following additive decomposition holds for any time series~$X$, which satisfies a weak integrability condition,
\begin{equation}\label{decomposition}
X(t) = E(X)(t) + X_{\tiny{\rm fluctuation}}(t)
\end{equation}
where
\begin{itemize}
\item the \emph{mean}, or \emph{trend}, $E(X)$ is ``quite smooth,'' 
\item $X_{\tiny{\rm fluctuation}}$ is quickly fluctuating.
\end{itemize}
Decomposition \eqref{decomposition} is unique up to an additive
infinitesimal.

\subsubsection{Forecasting}
Start with the first degree polynomial time function $p_{1} (\tau)
= a_0 + a_1 \tau$, $\tau \geqslant 0$, $a_0, a_1 \in \mathbb{R}$. Rewrite
thanks to classic operational calculus (see, \textit{e.g.} \cite{yosida}) with respect to the variable $\tau$, $p_1$ as $P_1 = \frac{a_0}{s} +
\frac{a_1}{s^2}$. Multiply both sides by $s^2$:
\begin{equation}\label{5}
s^2 P_1 = a_0 s + a_1
\end{equation}
Take the derivative of both sides with respect to $s$, which
corresponds in the time domain to the multiplication by $- \tau$:
\begin{equation}\label{6}
s^2 \frac{d P_1}{ds} + 2s P_1 = a_0
\end{equation}
The coefficients $a_0, a_1$ are obtained via the triangular system
of equations (\ref{5})-(\ref{6}). We get rid of the time
derivatives, \textit{i.e.}, of $s P_1$, $s^2 P_1$, and $s^2 \frac{d
P_1}{ds}$, by multiplying both sides of Equations
(\ref{5})-(\ref{6}) by $s^{ - n}$, $n \geqslant 2$. The corresponding
iterated time integrals are low pass filters which attenuate the
corrupting noises. A quite short time window is sufficient for
obtaining accurate values of $a_0$, $a_1$. Note that estimating $a_0$ and $a_1$ yields respectively the mean and the derivative\footnote{See \cite{mboup} and \cite{othmane1,othmane2} for more details.}.

Set the following forecast $X_{\text{forecast}}(t + \Delta T)$, where $\Delta T > 0$ is not too ``large,''
\begin{equation}\label{forecast}
X_{\text{forecast}}(t + \Delta T) = E(X)(t) + \left[\frac{d E(X)(t)}{dt}\right]_e \Delta T
\end{equation} where $E(X)(t)$ and $\left[\frac{d E(X)(t)}{dt}\right]_e$ are estimated like $a_0$ and $a_1$ above. Let us stress that what we predict is the mean and not the quick fluctuations\footnote{Compare with classical approaches to time series, like, \textit{e.g.}, in the nice textbook \cite{melard}.}.

\subsection{Application to Model-Free Control}
The popular ultra-local model \eqref{1} is replaced by Equation~\eqref{2}. The data-driven integral formula for estimating $F$ stems at 
once from \cite{mfc1}:
{\small
\begin{equation}\label{io}
F_{\rm est} (t) =  \frac{6}{\tau^{3}} \int_{t-\tau}^t \left( (t-2\sigma)\delta q (\sigma) + \alpha \sigma(\tau - \sigma)\delta u(\sigma - h)\right)d\sigma
\end{equation}}
In practice  Formula \eqref{io} is replaced by a simple digital filter. View $F_{\rm est} (t)$ as a time series. Via Formula \eqref{forecast}, $F_{\rm est}(t + h)$ is calculated at once. Equation \eqref{2} gives the estimate $\widehat{\delta q} (t + h)$ of the output at time $t + h$. Replace Equation \eqref{2} by
\begin{equation}\label{2bis}
\frac{d}{dt} \widehat{\delta q }(t + h) = F_{\rm est}(t + h) + \alpha \delta u(t)   
\end{equation}
Introduce now the \emph{intelligent proportional feedback with delay}:
\begin{equation}\label{ipdelay}
 \delta u (t) = \frac{\dot{{\delta q^\star}} (t+ h) - F_{\rm est}(t + h) - K_P \widehat{\delta e} (t + h)}{\alpha} \end{equation}
 where 
 \begin{itemize}
 \item $\delta q^\star$ is a reference trajectory,
 \item $\delta e = \delta q - \delta q^\star$ is the tracking error, 
 \item $\widehat{\delta e} (t + h) = \widehat{\delta q} (t + h) - {\delta q^\star} (t + h)$.
 \end{itemize}
where
$$
\widehat{\delta q} (t + h) = \delta q(t) + \int_t^{t+h} F_{\text{est}} (\tau) d\tau + \alpha \int_{t-h}^t u(\tau) d\tau 
$$
 Combining Equations \eqref{2bis} and \eqref{ipdelay} yields $\frac{d}{dt}\widehat{\delta e} (t + h) + K_P \widehat{\delta e} (t + h) = 0$. If the estimates are ``good,'' stability is ensured with ${K_P} > 0$.
 \begin{remark}
 Formulae \eqref{forecast} and \eqref{io} permit to avoid the constructions of complex observers which are often encountered in other settings.   
 \end{remark}
 
\section{Computer experiments}\label{simu}

\begin{figure*}[!h]
\centering%
{\epsfig{figure=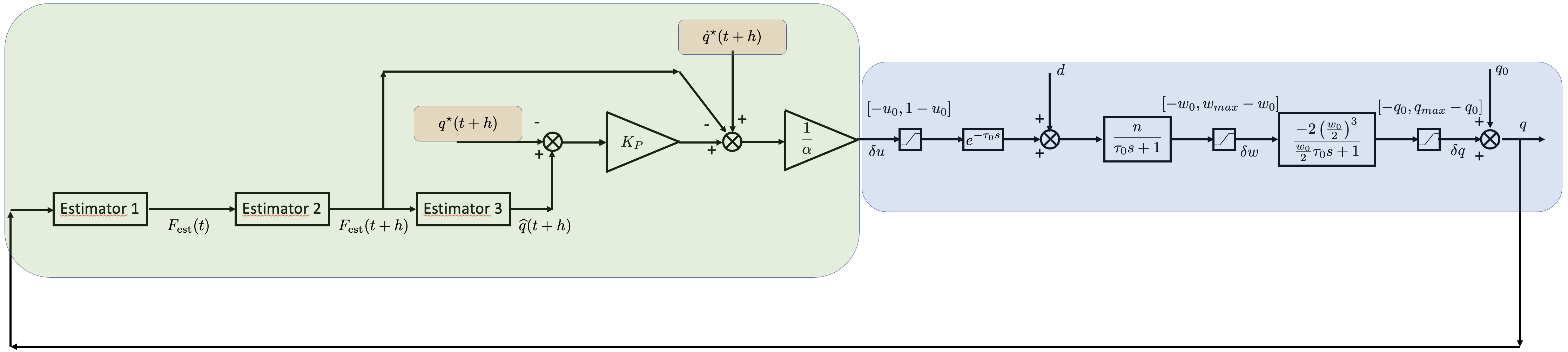,width=0.92\textwidth}}
\caption{Closed loop scheme}\label{CL}
\end{figure*}

\subsection{Transfer function}\label{transfer}
The variables $u(t)$, $0 \leqslant u \leqslant 1$, and $q(t)$, $0 \leqslant q \leqslant q_{\rm max}$, denote respectively the \emph{packet-loss ratio}\footnote{This terminology is borrowed from \cite{olu}. In most publications $\delta u$ is called \emph{drop-probability}.} and the \emph{queue-length} [packets]. Let $u_0$, $q_0$ be their numerical values at an operating point. Define the control variable $\delta u = u-u_0$ and the output variable $\delta q = q - q_0$. They are related by the time-invariant linear delay system defined by the transfer function (see comments and related references in \cite{olu}):
\begin{equation}\label{transf}
T(s) = - \frac{ (2N\frac{w_0}{2})^3 e^{- \tau_0 s}}{(\tau_0 s + 1)(\frac{w_0 \tau_0}{2} s + 1)}    
\end{equation}
where
\begin{itemize}
\item $w_0$ is the numerical value at the operating point of the length $w(t)$ of the TCP window,
\item $\tau_0$ is the numerical value at the operating point of the round trip time $\tau$,
\item $N$, which is assumed to be constant, is the number of TCP sessions,
\end{itemize}
The I/O system corresponding to the transfer function \eqref{transf}, which is sometimes called \cite{marquez} 
\emph{quasi-finite}\footnote{The output $\delta q$ is said \cite{marquez} to be {\em flat} or \emph{basic}.}, is only used for simulations purposes and not for feedback control. 

The numerical values of the parameters are in the Table~\ref{tb} (see \cite{olu}) below, where $T_p$ is the propagation delay:

\begin{table}[h!]
\begin{center}
\caption{Nominal system parameters}\label{tb}
\resizebox{0.48\textwidth}{!}{
\begin{tabular}{c|c|c|c|c|c|c|c|c|c}
Names &$w_{max}$&$q_{max}$&$q_0$& $N$&$c$& $T_p$&$\tau_0$&$w_0$&$u_0$\\\hline
Values& 131 & 800& 175&60&3750& 0.2 &$\frac{q_0}{c}+T_p$ &$\frac{c\tau_0}{N} $&$\frac{2}{w_0^2}$
\end{tabular}}
\end{center}
\end{table}

\subsection{Control and block diagram}\label{block}

Equations \eqref{2} and \eqref{ipdelay} define our control synthesis. Let us emphasize that in this model-free setting only the time delay $\tau_0$ is assumed to be known \textit{a priori}, \textit{i.e.}, independently of the input-output data. It yields the block diagram of Figure~\ref{CL}:
\begin{itemize}
\item The right blue part corresponds to the I/O system defined by Equation \eqref{transf}.
\item Estimators $1$, $2$ and $3$ in the left green part are respectively defined by Equations \eqref{io}, \eqref{forecast} and \eqref{2bis}.
\end{itemize}

\subsection{Computer experiments with the estimators}\label{estim}

\begin{figure*}[!h]
\centering%
\subfigure[\footnotesize $F_{\rm est} (t)$ (red, - -), $F_{\rm est} (t+\tau_0)$ (blue, -.-) and $F(t+\tau_0)$ (black, --)]
{\epsfig{figure=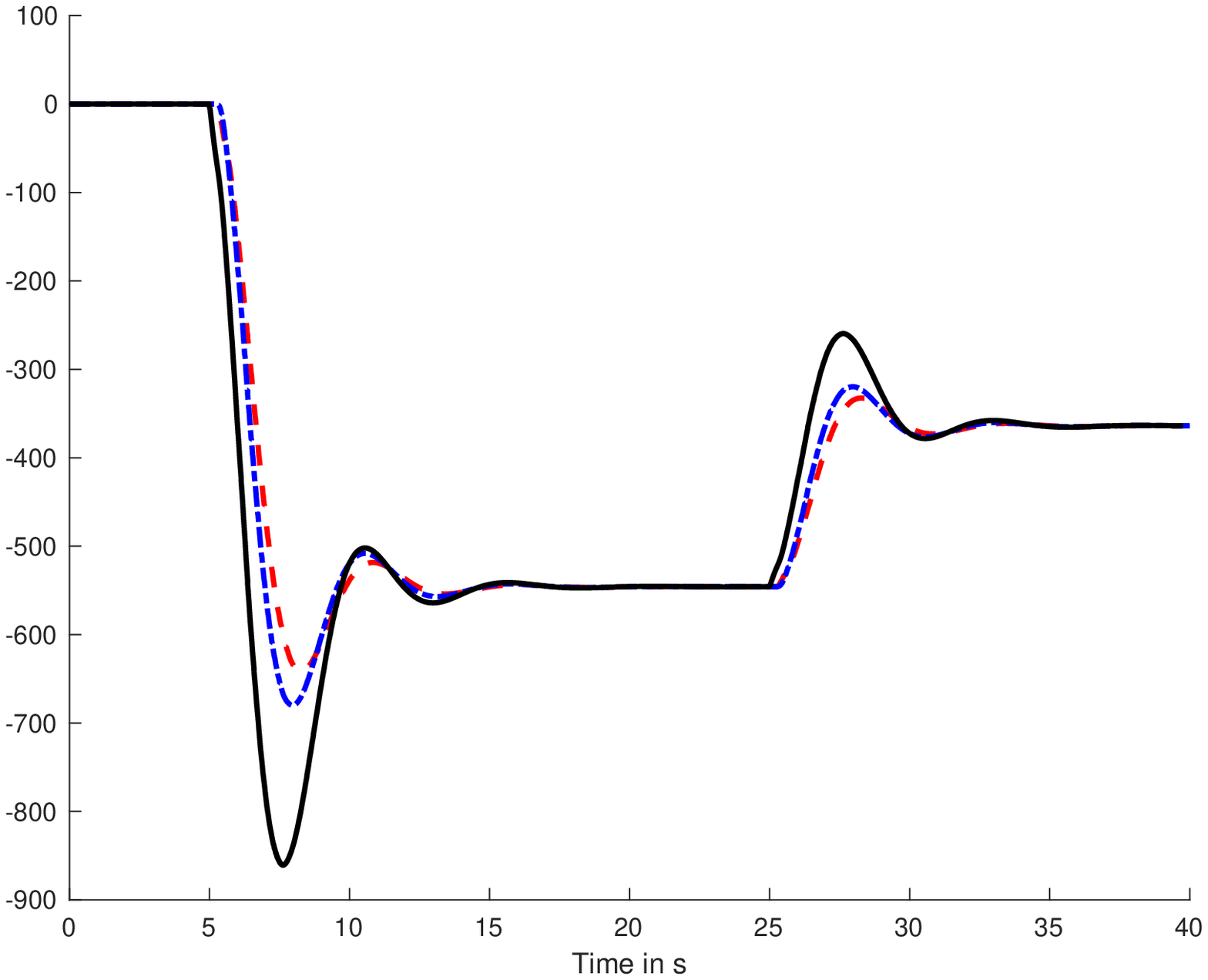,width=0.33\textwidth}}
\subfigure[\footnotesize Zoom on Figure \ref{E2}-(a)]
{\epsfig{figure=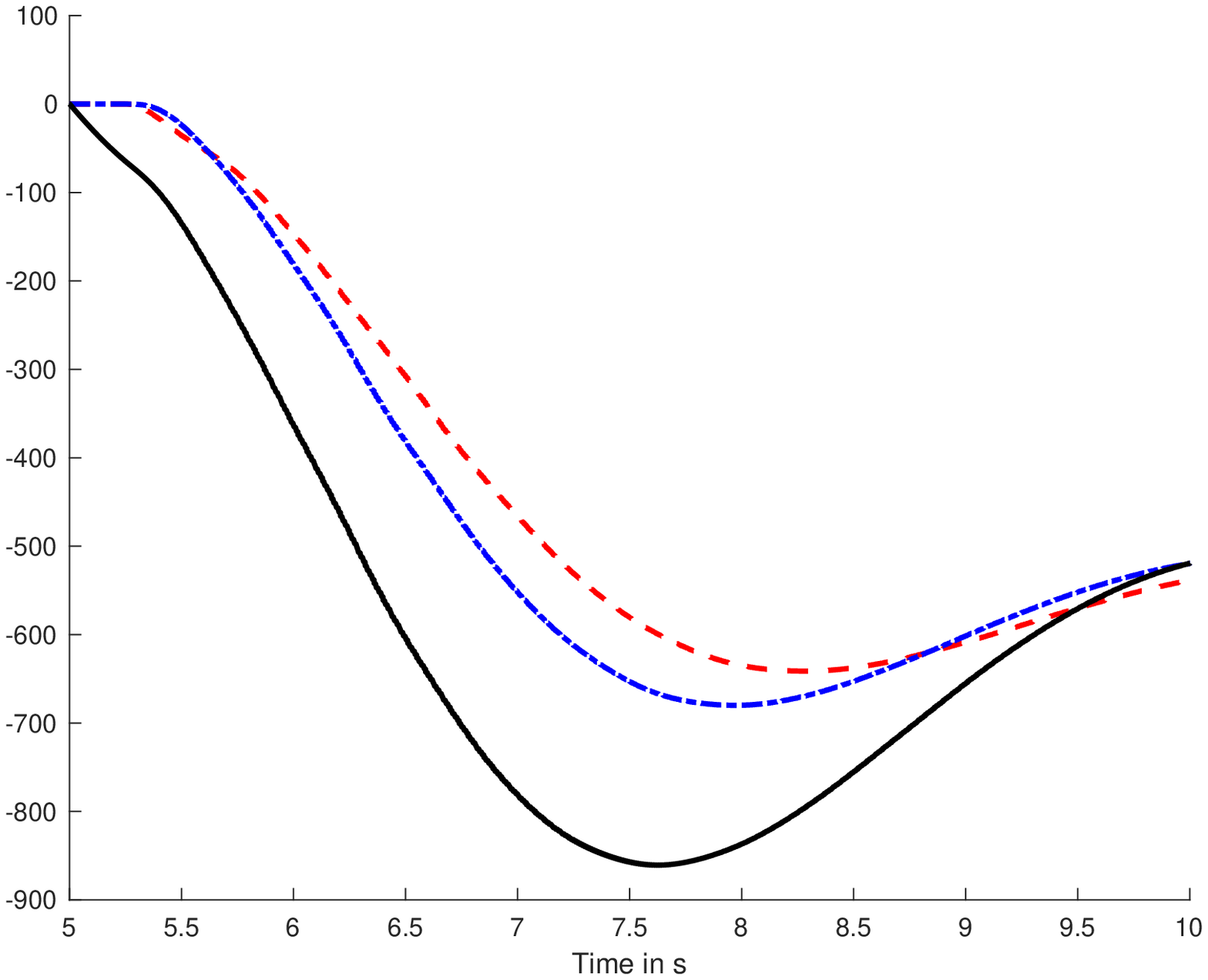,width=0.33\textwidth}}
\caption{Estimators 1 and 2}\label{E2}
\end{figure*}

\begin{figure*}[!h]
\centering%
\subfigure[\footnotesize $\delta q (t)$ (red, - -), $\widehat{\delta q} (t+\tau_0)$ (blue, -.-) and $\delta q(t+\tau_0)$ (black, --)]
{\epsfig{figure=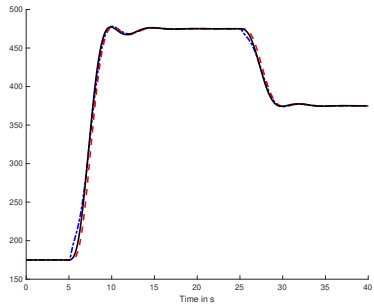,width=0.33\textwidth}}
\subfigure[\footnotesize Zoom on Figure \ref{E3}-(a)]
{\epsfig{figure=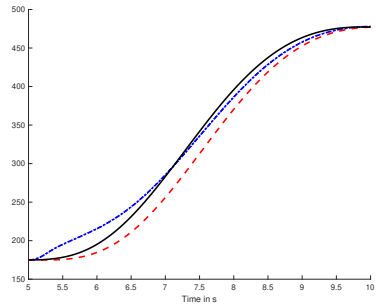,width=0.33\textwidth}}
\caption{Estimator 3}\label{E3}
\end{figure*}

For simplicity's sake there is no measurement noise in the computations displayed in Figures~\ref{E2} and~\ref{E3}. The excellent predictions in Figure \ref{E3}, which really matter, correct the mediocrity of the results in Figure \ref{E2}.

\subsection{Scenarios 
}\label{scenario}
Set in Equations \eqref{2bis} and \eqref{ipdelay} $\alpha = -10^5$, $K_P = -0. 5$. The duration of any scenario is $35$\,s. The sampling period is $10$\,ms.

\subsubsection{Nominal case} 
\begin{figure*}[!ht]
\centering%
\subfigure[\footnotesize Control]
{\epsfig{figure=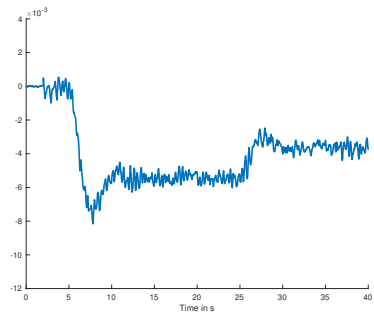,width=0.33\textwidth}}
\subfigure[\footnotesize Output (--) and output reference (- -)]
{\epsfig{figure=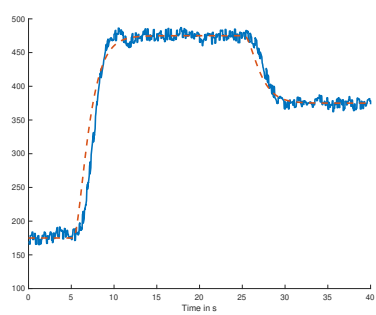,width=0.33\textwidth}}
\caption{With nominal round trip delay}\label{SN:1}
\end{figure*}

With respect to the values in Table~\ref{tb}, Figure~\ref{SN:1} shows excellent tracking performances.

\subsubsection{Round trip time and TCP sessions}
\begin{figure*}[!ht]
\centering%
\subfigure[\footnotesize Control]
{\epsfig{figure=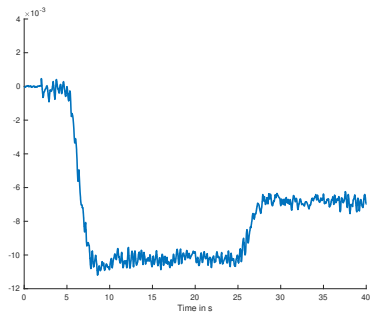,width=0.33\textwidth}}
\subfigure[\footnotesize Output (--) and output reference (- -)]
{\epsfig{figure=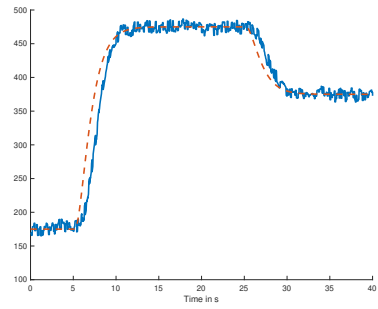,width=0.33\textwidth}}
\caption{Without delay}\label{S0}
\end{figure*}

\begin{figure*}[!ht]
\centering%
\subfigure[\footnotesize Control]
{\epsfig{figure=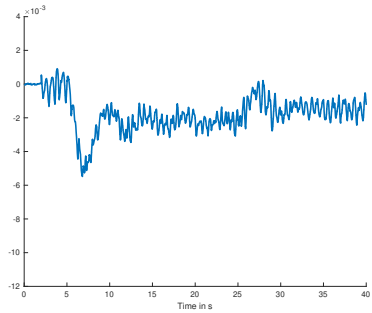,width=0.33\textwidth}}
\subfigure[\footnotesize Output (--) output reference (- -)]
{\epsfig{figure=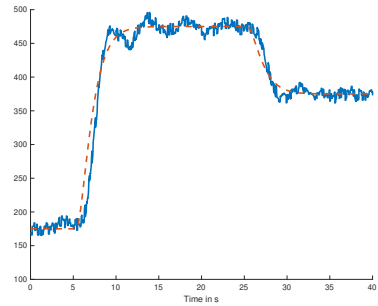,width=0.33\textwidth}}
\caption{With delay greater than the real round trip delay}\label{SI}
\end{figure*}

A poor evaluation of the round trip time $\tau_0$ leads to situations where, for smaller (resp. larger) delays, Figure \ref{S0} (resp. \ref{SI}) displays most satisfying performances. 
Similar results are obtained if we change the number $N$ of TCP sessions. Lack of space does not allow to present the corresponding Figures.

\subsubsection{Exogenous perturbations}
\begin{figure*}[!ht]
\centering%
\subfigure[\footnotesize Control]
{\epsfig{figure=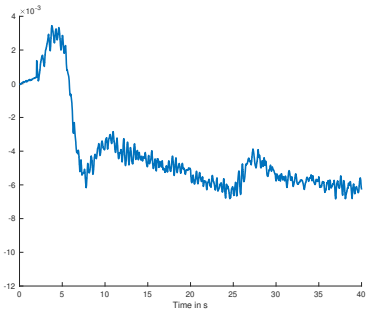,width=0.33\textwidth}}
\subfigure[\footnotesize Output (--) and output reference (- -)]
{\epsfig{figure=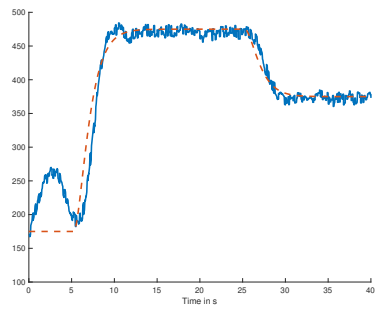,width=0.33\textwidth}}
%
\subfigure[\footnotesize Disturbance]
{\epsfig{figure=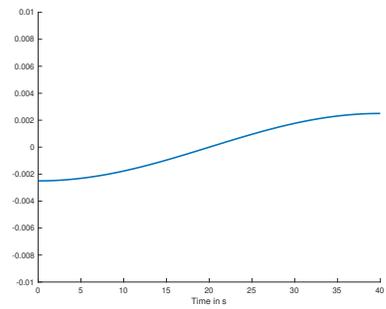,width=0.33\textwidth}}
\caption{First simulation with disturbance}\label{P1}
\end{figure*}

\begin{figure*}[!ht]
\centering%
\subfigure[\footnotesize Control]
{\epsfig{figure=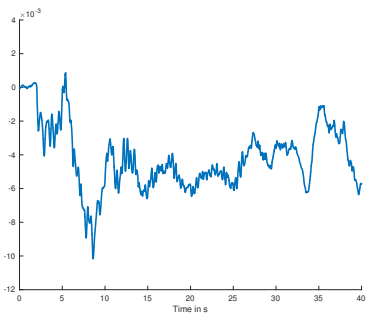,width=0.33\textwidth}}
\subfigure[\footnotesize Output (--) output reference (- -)]
{\epsfig{figure=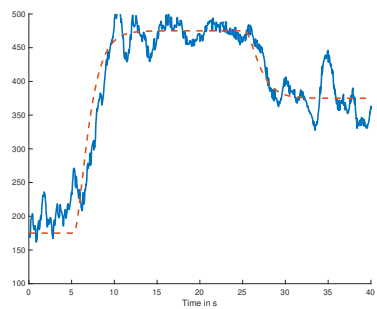,width=0.33\textwidth}}
%
\subfigure[\footnotesize Disturbance]
{\epsfig{figure=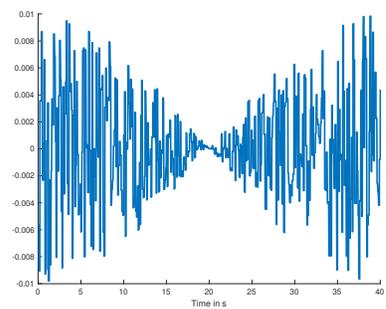,width=0.33\textwidth}}
\caption{Second simulation with disturbance}\label{P2}
\end{figure*}

The origins of such perturbations 
can be manifold. Let us mention here the presence of packets related to the \emph{User Datagram Protocol} (see, \textit{e.g.}, \cite{udp}), or \emph{UDP}. In order to carry on computer experiments, consider additive perturbations with respect to the control variable, which are either sinusoidal (see Figure \ref{P1}-(c)) or random (see Figure \ref{P2}-(c)) where it is expressed by the product $\mathcal{U}(-10^{-2},10^{-2}) \times \sin(\frac{\pi}{40}t+\frac{\pi}{2} )$ of a uniform noise and a sine wave. The performances do not deteriorate. 




\section{Conclusion}\label{conclu}
A future publication \cite{nl} will demonstrate, among other things, that keeping the delay frozen in the nonlinear modeling \cite{misra} (see also \cite{mounier}) implies that the other system variables, in particular, the packet-loss ratio, \textit{i.e.}, the control variable, are also kept constant. This fact casts some doubts on the works exploiting the linear approximation with a time-invariant delay, like in Formula \eqref{transf}.

According to the \textit{Cisco Visual Networking Index: Forecast and Methodology 2016-2021}, Internet video viewing is estimated to account for approximately $82\%$ of the Internet traffic in 2022. Besides maintaining the \emph{Quality of Service}, or \emph{QoS}, a \emph{Quality of Experience}, or \emph{QoE}, which indicates the user-side satisfaction, has also been investigated (see, \textit{e.g.}, \cite{qoe}). Several papers (see, \textit{e.g.}, \cite{huang,qin,sakamoto}) suggests the use of PID-like regulators for adjusting the bitrate. MFC should therefore be all the more efficient as no precise mathematical modeling seems to be available.


\end{document}